\documentclass[letterpaper, 10 pt, conference]{ieeeconf}  

\IEEEoverridecommandlockouts                              
\usepackage{graphics} 
\usepackage{epsfig} 
\usepackage{mathptmx} 
\usepackage{times} 
\usepackage{amsmath} 
\usepackage{amssymb}  
\usepackage{tcolorbox}
\usepackage{algorithm}
\usepackage{placeins}
\usepackage{booktabs}
\usepackage{tabularx}
\usepackage{makecell}
\usepackage{tikz}

\newcommand\ieeesubmissiontext{%
    \small\centering
    This work has been submitted to the IEEE for possible publication. \\ Copyright may be transferred without notice, after which this version may no longer be accessible.}

\newcommand\ieeesubmissionnotice{%
  \begin{tikzpicture}[remember picture,overlay]
    \node[anchor=south, yshift=10pt] at (current page.south)
      {\parbox{\dimexpr\textwidth-2\fboxsep\relax}{\ieeesubmissiontext}};
  \end{tikzpicture}}

\title{\LARGE \bf
Distributive Perimetral Queue Balancing Mechanisms: \\ Towards Equitable Urban Traffic Gating and Fair Perimeter Control
}

\author{
    Kevin Riehl, Lea Künstler, Ying-Chuan Ni, Anastasia Psarou, \\ Shaimaa K. El-Baklish, Anastasios Kouvelas, Michail Makridis 
    \thanks{This work was not supported by any organization}
    \thanks{All authors are with Traffic Engineering Group, Institute for Transportation Planning and Systems, ETH Zürich, Stefano Franscini Platz 3, 8053 Zürich, Switzerland. Psarou is also with Doctoral School of Exact and Natural Sciences, Jagiellonian University, Kraków, Poland. Corresponding Author: kriehl@ethz.ch.}
}

\begin{document}

\maketitle
\ieeesubmissionnotice
\thispagestyle{empty}
\pagestyle{empty}

\begin{abstract}
Perimeter control is an effective urban traffic management strategy that regulates inflow to congested urban regions using aggregate network dynamics. While existing approaches primarily optimize system-level efficiency, such as total travel time or network throughput, they often overlook equity considerations, leading to uneven delay distributions across entry points. 
This work integrates fairness objectives into perimeter control design through explicit queue balancing mechanisms.
A large-scale, microscopic case study of the Financial District in the San Francisco urban network is used to evaluate both performance and implementation challenges. 
The results demonstrate conventional perimeter control not only reduces total and internal delays but can also improve fairness metrics (Harsanyian, Rawlsian, Utilitarian, Egalitarian). 
Building on this observation, queue balancing strategies match conventional performance while yielding measurable fairness improvements, especially in heterogeneous demand scenarios, where congestion is unevenly distributed across entry points. 
The proposed framework contributes toward equitable control design for emerging intelligent transportation systems and higher user acceptance for those. 

Code \& Resources: 

https://github.com/DerKevinRiehl/fair\_perimeter\_control
\end{abstract}

\section{INTRODUCTION}

Urban road transportation networks across the globe are facing steadily worsening congestion, as growing travel demand continues to outpace the available infrastructure capacity. 
Congestion arises when the number of vehicles attempting to use a network exceeds its physical and operational limits, leading to reduced speeds, longer travel times, and increased environmental and economic costs~\cite{daganzo2007urban}. 
In response, intelligent transportation systems (ITS) leverage sensor technologies and real-time data to monitor traffic conditions and enable more efficient control strategies. 
In particular, signalised intersections with advanced traffic light control can increase the effective capacity of road networks, allowing more vehicles to utilize the same infrastructure~\cite{kouvelas2011a}. 
However, a centralized control approach for all intersections is an extremely challenging task due to the computational complexity and varying travel patterns. 
Consequently, attention has increasingly shifted toward network-wide traffic management strategies, such as perimeter control, which regulates the number of vehicles entering a network in order to maintain stable and efficient traffic conditions~\cite{haddad2012stability,geroliminis2012optimal}.



Early works on perimeter control, such as the two reservoir model~\cite{daganzo2007urban}, established the basis for regulating inflow to protected areas based on the total number of driving vehicles in the cell. 
Building on this, feedback-based gating using real-time measurements and PI controllers~\cite{keyvan2012exploiting} for single cells, optimal model-predictive control~\cite{geroliminis2012optimal}, linear-quadratic integral feedback regulators for multiple cells~\cite{aboudolas2013perimeter}, robust perimeter control~\cite{haddad2014robust}, adaptive, data-driven perimeter control~\cite{haddad2020adaptive,kouvelas2017enhancing}, combinations with signalised intersection management~\cite{tsitsokas2023two,liu2024n}, time-varying cell borders~\cite{li2021perimeter}, coloured Petri nets~\cite{fu2021perimeter}, percolation-based approaches~\cite{hamedmoghadam2022percolation}, and machine-learning based approaches~\cite{zhou2023improving} have been proposed to keep the network near its optimal state, aiming to maximize throughput and avoid over-saturation, even in heterogeneous urban areas.

Most of these contributions rely on macroscopic traffic flow representations and focus on maximising efficiency metrics such as network throughput or minimising total travel time.
While macroscopic models provide analytical tractability and theoretical insight, they abstract from critical microscopic phenomena such as queue spill-backs, and gridlock propagation. 
These phenomena become particularly relevant in real-world implementations, where spatially heterogeneous demand and intersection-level constraints shape traffic dynamics. 
Recent work therefore increasingly evaluates perimeter control strategies using high-resolution microsimulation environments~\cite{zhou2024evaluating,kampitakis2025decentralized}. 
However, such microscopic settings reveal additional complexities that are not captured by aggregate control formulations.
To date, only one city has implemented a preliminary approach in practice (Zürich, Switzerland)~\cite{ambuhl2023understanding,ortigosa2015study}.


Furthermore, existing perimeter control literature is limited by its predominant focus on efficiency~\cite{riehl2024towards}. 
In practice, purely efficiency-driven gating can generate highly uneven queue lengths at the perimeter, producing excessive waiting times for specific origins while others experience relatively smooth access.
Such inequities are not merely ethical concerns but also system-level risks~\cite{riehl2024quantitative}. 
Traffic control strategies perceived as unfair may face political resistance, reduced compliance, or behavioural adaptations such as re-routing, signal violations, or strategic departure-time shifts, potentially undermining both safety and efficiency gains~\cite{moshahedi2023alpha,keyvan2021optimizing}. 
For emerging ITS solutions to achieve long-term acceptance, efficiency must therefore be complemented by explicit fairness considerations~\cite{riehl2024towards}.


Existing attempts to incorporate fairness into perimeter control include $\alpha$-fair formulations~\cite{moshahedi2023alpha} -- enabling the trade-off between efficiency and fairness through a tunable parameter -- and queue-balancing approaches~\cite{keyvan2021optimizing} -- demonstrating improvement of traffic conditions also outside the protected area. 
While these studies demonstrate the feasibility of distributive considerations, they are either tied to specific fairness ideologies or rely on optimisation frameworks that may be difficult to implement in large-scale, real-time microscopic environments. 
Moreover, fairness in traffic systems is inherently multidimensional and benefits from ideology-agnostic, systematic evaluation~\cite{riehl2024quantitative}.


This paper proposes a \emph{distributive, perimetral queue balancing mechanism} as an extension to classical feedback-based gating control. 
Instead of assigning identical green ratios to all boundary intersections of a zone, the proposed framework redistributes the available inflow budget according to local queue conditions while preserving the aggregate inflow determined by the zone-level controller. 
Doing so, two allocative principles are investigated, incorporating proportional (Aristotelian) and max-min (Rawlsian) fairness ideology.
The proposed mechanism is computationally lightweight, compatible with microscopic traffic simulation and real-time signal control, and directly applicable to multi-zone gating architectures.

The large-scale microsimulation case study of San Francisco's Financial District -- comprising three interacting zones and 46 signalised intersections -- demonstrates that conventional perimeter control can already improve certain fairness metrics by stabilising network conditions. 
Furthermore, particularly under heterogeneous demand scenarios, the proposed queue balancing mechanisms yields additional and measurable equity improvements without sacrificing overall efficiency, such as queue length and delay homogenization.

By explicitly integrating distributive principles into feedback-based gating, this work contributes toward equitable control design for intelligent transportation systems and advances the discussion from purely efficiency-oriented optimisation toward socially robust urban traffic management.


\FloatBarrier
\begin{figure} [h!]
    \centering
    \includegraphics[width=\linewidth]{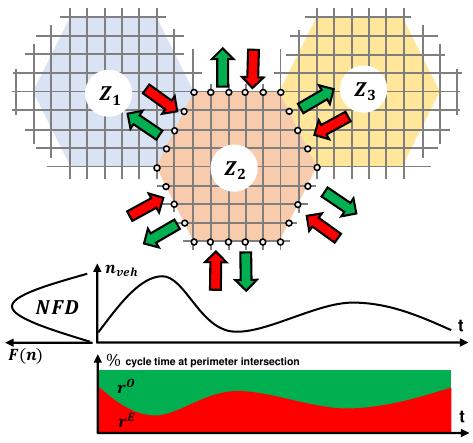}
    \caption{\textbf{Perimeter Control \& Signal Plans.}}
    \label{fig:perimeter_control}
\end{figure}

\section{PROBLEM FORMULATION}

In this section, the problem of perimeter control (Fig.~\ref{fig:perimeter_control}), and the notation are specified. Then, the feedback-based gating algorithm is introduced, as the proposed queue balancing mechanism bases on it.

\subsection{Definitions \& Notation}

An urban road network can be partitioned into certain zones / regions $Z_i \in \mathcal{Z}$.
During their daily commute inside cities, road users often start from many different origin zones outside the city centre and drive towards a zone of interest (such as the city centre, central zone).  
Each region $Z_i$ is characterised by a network fundamental diagram (NFD) which describes the relationship between the vehicle accumulation (number of vehicles) $n_i$ in the zone $Z_i$ and the vehicle flow $NFD = F_i(n_i)$ (measured in veh/h) that can be achieved.
The NFD typically follows a concave, u-shaped form, where the flow $F_i(n_i)$ increases as the accumulation $n_i$ increases if it is smaller than an optimal value $n_i^*$, and the flow is maximised at its capacity $C_i$ at the optimal accumulation, meaning $F_i(n_i^*) = C_i$; due to congestion that arises in a too crowded zone, the flow decreases for higher accumulations.
The goal of perimeter control is to limit the number of vehicles in a zone to achieve congestion-free, flow-maximising traffic conditions in the road network of that zone.
This is achieved by gating the signalised intersections $j \in\mathcal{J}_i$ at the perimeter (border) of that zone $Z_i$, and dynamically adapting the rate at which incoming roads get right-of-way into the zone $r_j^E \in [0,1]$ (respectively out the zone $r_j^O \in [0,1]$), where the cycle duration $t_c$ equals the sum of incoming and outgoing phases in the intersection's signal plan: $t_{c,j} = t_{E,j} + t_{O,j}$, with $t_{E,j} = r_j^E\times t_{c,j}$, $t_{o,j} = r_j^O \times t_{c,j}$, and $1=r_j^E+r_j^O$.
Once the number of vehicles surpasses sustainable levels beyond the capacity ($n_i > n_i^*$) the incoming rate $r_j^E$ is reduced, and the outgoing rate $r_j^O$ is increased. 

The control problem of perimeter control (for a single zone) is to adjust rates $r_j^E \; \forall \; j \in \mathcal{J}_i$ dynamically according to traffic conditions in that zone $n_i$, to achieve certain goals, such as maximising the flow, reflecting transportation system efficiency.

\subsection{Feedback-based Gating Algorithm}

The feedback-based gating algorithm (based on~\cite{keyvan2012exploiting}) is one of the simplest, time-discrete feedback perimeter controller.
It assumes the control of one zone $Z_i$ with occupancy $n_i$ and NFD-characteristic optimal number of vehicles $n_i^*$, and all intersections $j \in \mathcal{J}_i$ sharing the same, fixed cycle duration:
\begin{equation}
    t_{c,j} = t_c \; \forall \; j \in \mathcal{J}_i
\end{equation}
Furthermore, it assumes that all intersections at the perimeter share the same rates:
\begin{equation}\label{condition}
    r_j^O = r_i^O \; \forall \; j \in \mathcal{J}_i \;\;\; \text{and} \;\;\; r_j^E = r_i^E \; \forall \; j \in \mathcal{J}_i.
\end{equation}
It reduces the control problem to finding an optimal $r_i^E$ as the fixed-cycle conservation condition holds:
\begin{equation}
    r_i^O + r_i^E = 1
\end{equation}
At discrete time steps $k$ (after completing a intersection cycle duration $t_c$) the rate $r_i^E[k]$ is determined with a PI-control law as follows:
\begin{equation}
    r_i^E[k] = r_i^E[k-1] + K_P (n_i^T-n_i[k]) + K_I (n_i[k]-n_i[k-1])
\end{equation}
where $K_P$ and $K_I$ are control parameters that represent gain factors, and $n_i^T$ represents a target zone accumulation, which in practice can be set close to $n_i^* > n_i^T$.

The rate $r_i^E[k]$ is further bounded by a minimum rate $r_{min}^E$ to ensure a minimum time for vehicle inflow to a zone in order to avoid excessive queue formation and gridlock formation.

\subsection{Extension: Multi-Zone Gating Algorithm}

Two neighbouring zones $Z_a$ and $Z_b$ share a common border at their perimeter, including a set of intersections that are related to both zones: $j \in \mathcal{J}_a \cap \mathcal{J}_b$.
When both zones apply a specific feedback-based gating algorithm, then the rates $r_j^E$ for those intersections are set as a weighted average of those proposed by the single zone control laws $r_a^E$ and $r_b^E$:
\begin{equation}
r_j^E =
\begin{cases}
    r_a^E & \text{if } j \in \mathcal{J}_a \wedge j \notin \mathcal{J}_b \\
    r_b^E & \text{if } j \notin \mathcal{J}_a \wedge j \in \mathcal{J}_b \\
    \dfrac{r_a^E}{r_a^E + r_b^E} & \text{if } j \in \mathcal{J}_a \cap \mathcal{J}_b
\end{cases}
\end{equation}

In the case of multiple, gated zones, therefore, equation (\ref{condition}) does not hold for all intersection $j$, but just for those that are at a section of the perimeter which is intersection-free with other zones.


\begin{figure} [!h]
    \centering
    \includegraphics[width=\linewidth]{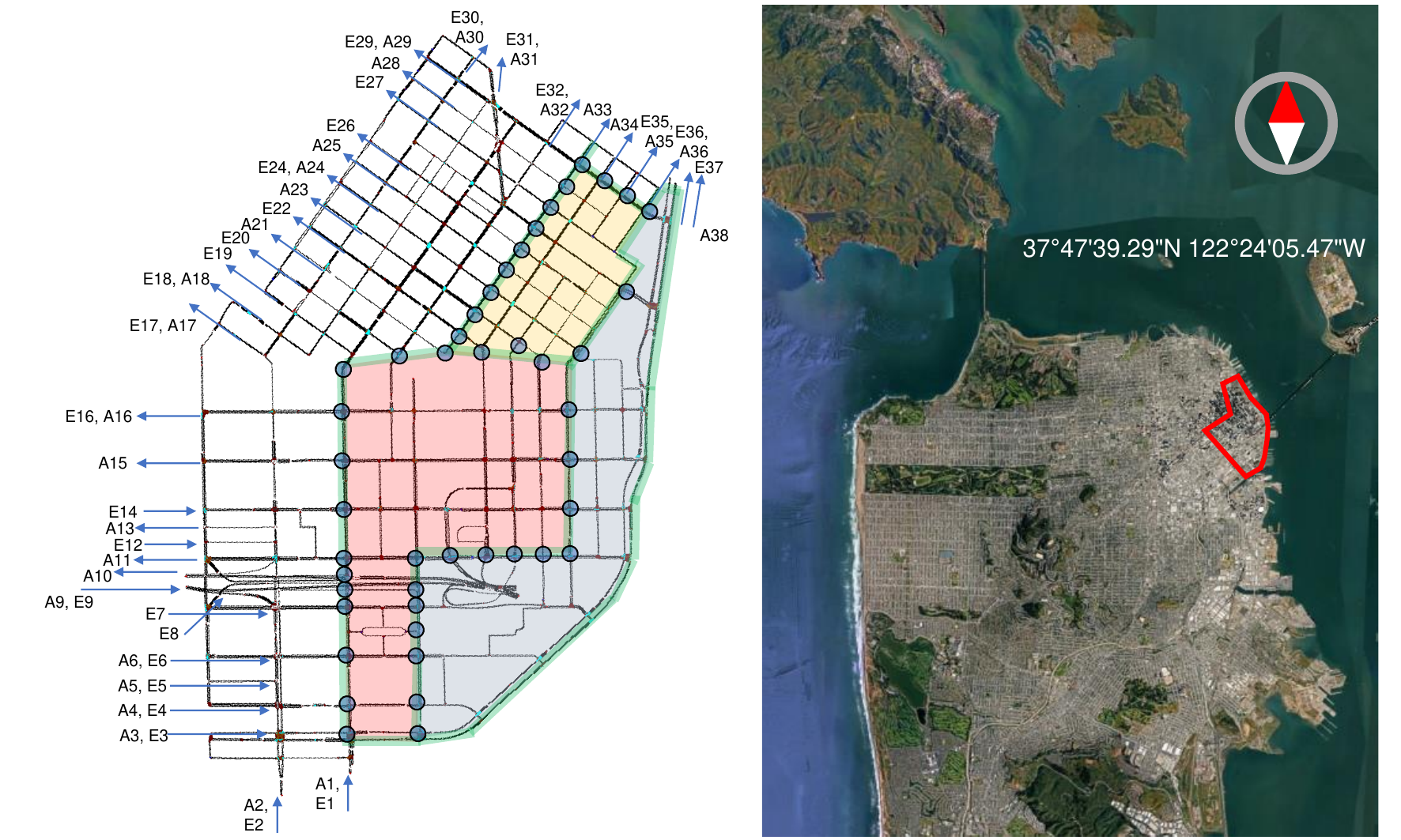}
    \caption{\textbf{Case Study: Three zones in San Francisco's Financial District.}}
    \label{fig:case_study}
\end{figure}

\FloatBarrier

\section{METHODS}

\subsection{Distributive, Perimetral Queue Balancing Mechanism}
The feedback-based gating algorithm determines an incoming rate $r_i^E$ for each zone $Z_i$. 
Doing so, it provides an available budget, the total admissible incoming green ratio $R^{\text{avail}}_i$, for all intersections of that zone:
\begin{equation}
    R^{\text{avail}}_i = \sum_{j \in \mathcal{J}_i} r_i^E
\end{equation}

The idea of the queue balancing mechanism, is that each intersection $j$ of the perimeter of $\mathcal{J}_i$ (zone $Z_i$) computes a requested incoming green ratio $\hat r_j^E \in [r_{min}^E,r_{max}^E]$ based on its local queue conditions (queue length).
This implies, that an intersection with shorter queues, reduces its rate $r_j^E < r_i^E$ and provides that green time to other intersections with longer queues, so that in total the same in-flow of vehicles is guaranteed but local queue length situations can be taken into account.

The requests are determined based on a distributive allocation rule, to resolve the conflicts between intersections.
Two distributive allocation rules are discussed in the following: (i) proportional queue balancing, and (ii) max-min queue balancing.

The sum of request cannot exceed the available budget:
\begin{equation}
    \sum_{j \in \mathcal{J}_i} \hat r_j^E \leq R^{\text{avail}},
\end{equation}

The determined requests are then granted to the intersections:
\begin{equation}
    r_j^E = \hat r_j^E.
\end{equation}

The aforementioned multi-zone extensions to the gating algorithm apply afterwards similarly.

\subsection{Proportional Queue Balancing Mechanism}

Each intersection $j$ raises a request $\hat r_j^E$ given its queue length $l_j$ representing the number of vehicles on all the intersection's approaches that would drive into the zone $Z_i$. 
The proportional distributive allocation rule determines the requests proportional to all the requests:
\begin{equation}
    \hat r_j^E = \frac{l_j}{\sum_{j \in \mathcal{J}_i} l_j}
\end{equation}

This mechanism preserves the relative proportions of the requests while
ensuring feasibility of the global gating constraint.
It can be interpreted as a weighted sharing mechanism in which larger
queues (reflected by larger $\hat r_j^E$) receive proportionally larger
allocations.
This reflects Aristotelian (proportional)~\cite{goppel2016handbuch}, and Prioritarian~\cite{riehl2024quantitative} fairness ideology, as those with greater needs (longer queues) receive at the cost of those with fewer needs~\cite{riehl2024towards}.

\subsection{Max-Min Queue Balancing Mechanism}

As an alternative to proportional allocation, a max-min fair distribution
mechanism is implemented.
The available time budget is distributed according to
a max-min fairness principle.
The iterative procedure is described in Algorithm~\ref{alg:1}.

\begin{algorithm}[!h]
    \caption{Iterative Max-Min Allocation Mechanism}
    \label{alg:1}
    \scriptsize
    \centering
    \begin{tcolorbox}[width=0.98\linewidth]
    \begin{enumerate}
        \item Sort the requests $\hat r_j^E$ in increasing order.
        \item For each remaining $j \in \mathcal{J}_i^{rem}$:
        \begin{enumerate}
            \item Compute equal share of remaining budget:
            \begin{equation}
                \rho = \frac{ \sum_{j \in \mathcal{J}_i^{rem}} l_j }{ \| \mathcal{J}_i^{rem} \|},
            \end{equation}
            \item All intersections whose request satisfies
            $\hat r_j^E = l_j \leq \rho$ receive their full request, and are removed from the set of remaining intersections.
            \item If no such intersection exists, all remaining intersections
            receive the equal share $\rho$, and the procedure terminates.
        \end{enumerate}
    \end{enumerate}
    \end{tcolorbox}
\end{algorithm}


This mechanism ensures max--min fairness, meaning that no intersection
can increase its allocation without decreasing the allocation of another
intersection with an equal or smaller granted share.
Compared to proportional allocation, this rule favours intersections
with large requests and prevents starvation under highly asymmetric
queue conditions.
This reflects Rawlsian~\cite{goppel2016handbuch}, and Max-Min~\cite{riehl2024quantitative} fairness ideology, as a focus is laid on those worst off~\cite{riehl2024towards}.

\subsection{Simulation Case Study}

To demonstrate the potential efficiency and equity improvements of the proposed method, we developed a microsimulation-based, demand-calibrated case study of the financial district of San Francisco (USA), that attracts significant traffic during daily commute from around the Bay-Area.
The network comprises three zones (similar to~\cite{aboudolas2013perimeter}), and 46 signalised intersections, as shown in Fig.~\ref{fig:case_study}.
Each perimetral intersection features lane-area detectors to measure queue lengths and induction loops at the gates to count vehicles within zones, operating with two distinct movement phases—either inflow to or outflow from the cell—to avoid turning conflicts.
In total 284 induction loop detectors and 131 lane area detectors are deployed.
The demand model is calibrated based on the mesoscopic simulation models from~\cite{ambuhl2023understanding}, and covers more than 125,000 trips from 38 origins to 37 destinations across $24~h$ of simulation.
The simulation model is implemented in SUMO~\cite{sumo2018} and Python, and repeated for 10 different random seeds (demand models).
Optimal parameters for the feedback-based gating algorithm have been identified as follows: region 1 \{ $n^T=45$, $K_P=0.007$, $K_I=0.001$ \}, region 2 \{ $n^T=290$, $K_P=0.007$, $K_I=0.001$ \}, region 3 \{ $n^T=240$, $K_P=0.007$, $K_I=0.001$ \}. The signal cycle duration is set to $t_c=90~s$, and rates are bounded to $r_{min}^E=20~\%$ and $r_{max}^E=50~\%$.


\section{RESULTS}

\begin{figure}[!ht]
    \centering
    \includegraphics[width=0.85\linewidth]{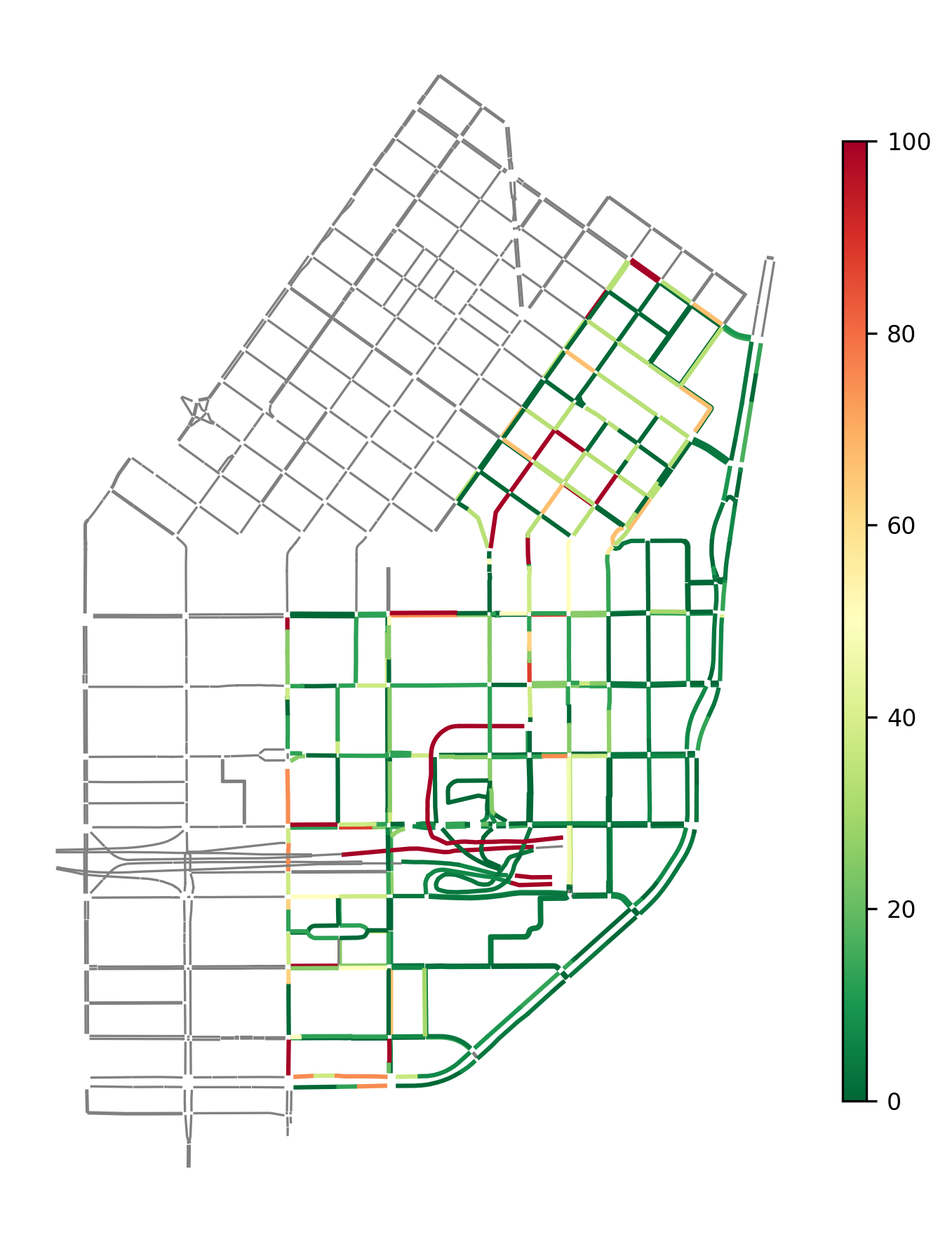}
    \caption{\textbf{Vehicle Accumulation [\%] at 09:00 (Uncontrolled).}}
    \label{fig:placeholder}
\end{figure}

In this section, we evaluate efficiency and fairness gains using multiple metrics for three scenarios: (i) uncontrolled, (ii) controlled with the feedback-based gating algorithm, and (iii) controlled with perimetral queue balancing. Fig.~\ref{fig:placeholder} illustrates congestion in the uncontrolled case, where vehicle accumulation exceeds capacity during morning peak at perimetral intersections in Zones 1 and 2, and on feeders for Zone 3 (bridge to Financial District from North East).

\subsection{Efficiency Gains with Gating Algorithm}

\begin{figure*}
    \centering
    \includegraphics[width=\linewidth]{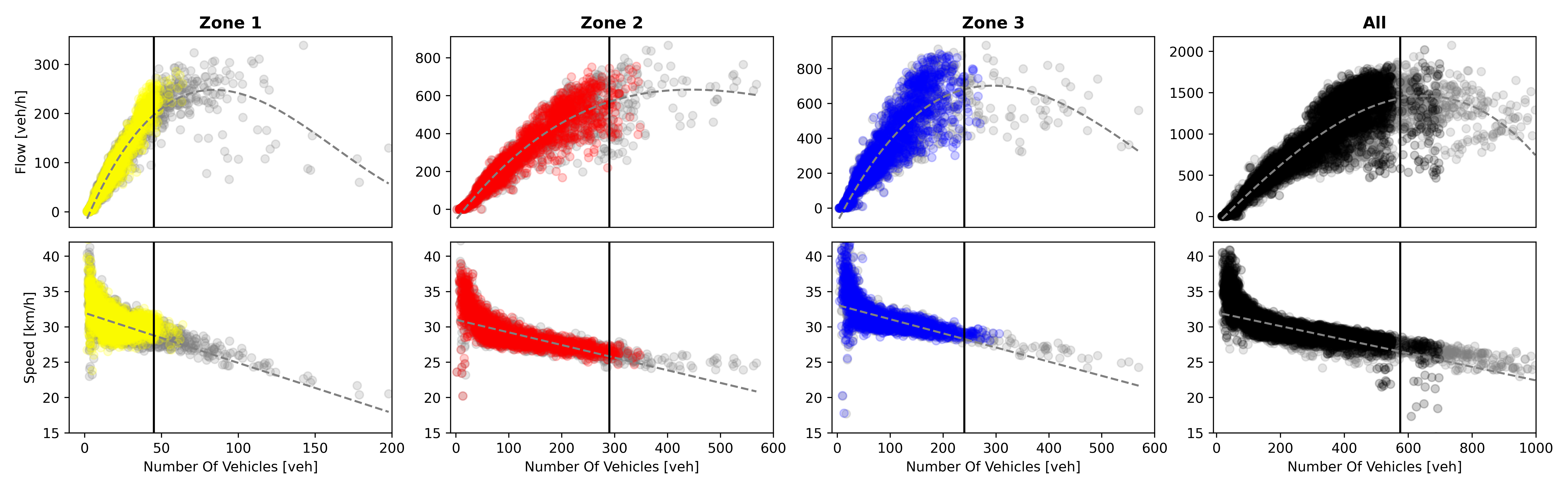}
    \includegraphics[width=\linewidth]{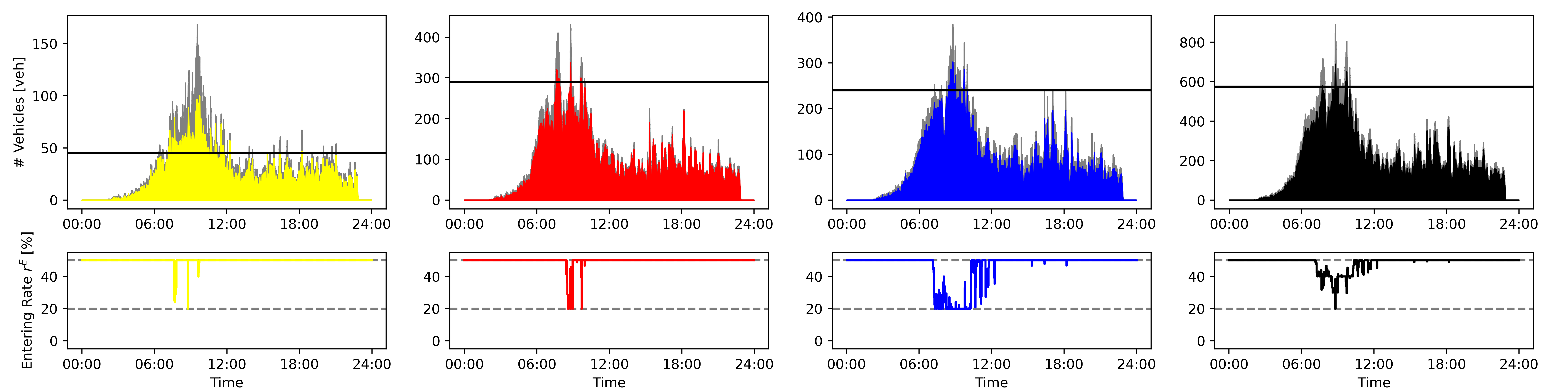}
    \includegraphics[width=\linewidth]{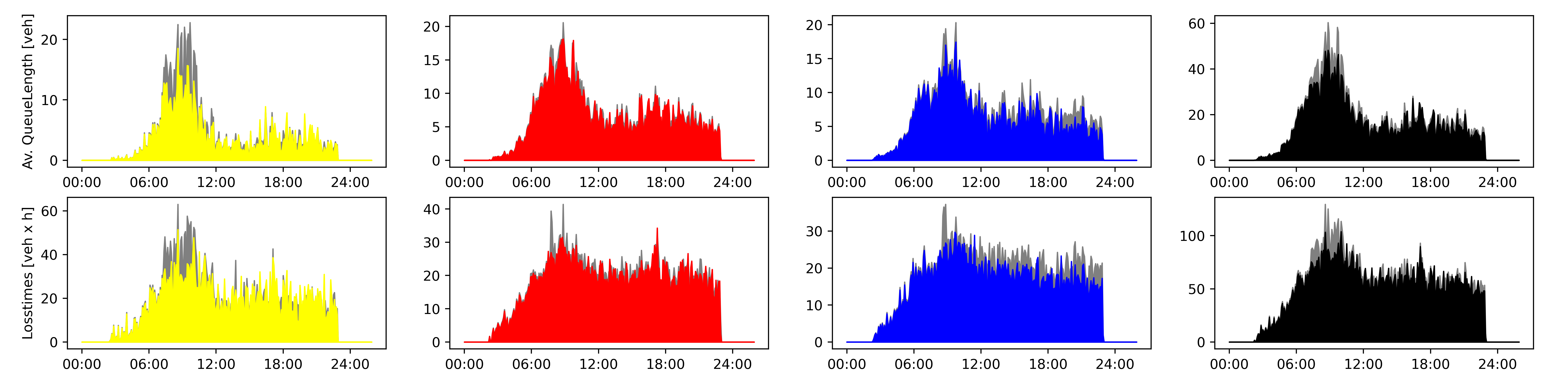}
    \caption{\textbf{Perimeter Control Implications on Fundamentals, Queue Lengths and Loss-Times.} Figures shown for different zones (coloured equals with perimeter control, gray uncontrolled) for one scenario qualitatively (similar results for different random seeds); all forms of perimeter control (Gating, MaxMin- and Proportional Queue Balancing yield similar results).
    Black solid lines in first two rows represent the zone's optimal number of vehicles $n^*$.
    Black dashed lines in third row represent minimum and maximum entering rates $r^E_{min}$ and $r^E_{max}$.
    }
    \label{fig:nfd}
\end{figure*}

Fig.~\ref{fig:nfd} (rows 1\&2) shows flow and speed versus vehicle accumulation (NFDs) for the three zones (yellow, red, blue) and their aggregate (black), with gray dots indicating the uncontrolled scenario. 
The NFDs indicate that the feedback-based gating algorithm stabilizes vehicle accumulation at or below capacity, increasing average flow by $\sim 13\%$ (662.05 vs. 581.35 veh/h uncontrolled) and speed by $\sim 16\%$ (30.06 vs. 25.82 km/h uncontrolled). 
Fig.~\ref{fig:nfd} (row 3) reports vehicle accumulation by zone, revealing that peak accumulations in Zones 1 and 2 are effectively reduced relative to the uncontrolled case, thereby maintaining operation at or below capacity. 
Fig.~\ref{fig:nfd} (row 4) shows the entering rates $r^E_i$ (and aggregate), where during the morning peak (08:30-12:00) the controllers frequently reduce the rate from $r^E_{\max}$ toward $r^E_{\min}$, particularly in Zone 3. 
Fig.~\ref{fig:nfd} (rows 5\&6) displays the impact on perimetral queue lengths and individual loss times, which are markedly reduced during peak periods across all zones. 
Comparable improvements are obtained with the perimetral queue balancing mechanisms.

\begin{table*}[htbp]
    \centering
    \scriptsize
    \caption{\textbf{Queue and LossTime Statistics At Intersections by Zone and Controller  (across scenarios / random seeds).}}
    \begin{tabularx}{\textwidth}{l l *{8}{>{\raggedleft\arraybackslash}X}}
        \toprule
        \textbf{Zone} & \textbf{Controller} & \makecell[c]{\textbf{Queues}\\\textbf{Mean}} & \makecell[c]{\textbf{Queues}\\\textbf{Max}} & \makecell[c]{\textbf{Queues}\\\textbf{Sum}} & \makecell[c]{\textbf{Queues}\\\textbf{Std}} & \makecell[c]{\textbf{LossTimes}\\\textbf{Mean}} & \makecell[c]{\textbf{LossTimes}\\\textbf{Max}} & \makecell[c]{\textbf{LossTimes}\\\textbf{Sum}} & \makecell[c]{\textbf{LossTimes}\\\textbf{Std}} \\
        \midrule
        1 & None & 3.28 (0.18) & 19.30 (1.50) & 72.06 (4.01) & 5.20 (0.35) & 2592 (93) & 11264 (542) & 57037 (2064) & 3302 (105) \\
        1 & Gating   & 3.89 (0.30) & 20.36 (1.76) & 85.54 (6.68) & 5.71 (0.45) & 2728 (78) & 11504 (376) & 60033 (3922) & 3249 (135) \\
        1 & MaxM-QB   & 3.22 (0.18) & \textbf{17.83 (1.65)} & 70.75 (3.87) & 5.17 (0.36) & 2564 (91) & \textbf{10982 (397)} & 56420 (2002) & 3278 (115) \\
        1 & Prop-QB   & \textbf{3.15 (0.25)} & 19.25 (1.48) & \textbf{69.16 (4.25)} & \textbf{4.89 (0.43)} & \textbf{2483 (89)} & 11414 (428) & \textbf{54626 (1897)} & \textbf{3143 (124)} \\
        \midrule
        2 & None & 5.06 (0.37) & 29.72 (1.01) & 288.50 (21.28) & 7.00 (0.37) & 923 (32) & 4265 (363) & 52632 (1871) & 983 (50) \\
        2 & Gating   & 4.98 (0.37) & 29.70 (1.15) & 283.82 (20.88) & 6.95 (0.39) & 907 (32) & 4349 (318) & 51703 (1855) & 967 (53) \\
        2 & MaxM-QB   & 4.97 (0.36) & \textbf{28.83 (1.04)} & 283.26 (20.41) & 6.93 (0.37) & 903 (31) & \textbf{4308 (336)} & 51512 (1785) & 970 (53) \\
        2 & Prop-QB   & \textbf{4.96 (0.38)} & 29.64 (1.16) & \textbf{282.84 (20.6)} & \textbf{6.91 (0.41)} & \textbf{902 (32)} & 4312 (354) & \textbf{51487 (1801)} & \textbf{966 (52)} \\
        \midrule
        3 & None & 4.77 (0.37) & 28.32 (1.72) & 243.23 (18.62) & 6.50 (0.48) & 946 (21) & 4153 (272) & 48255 (1103) & 952 (37) \\
        3 & Gating   & 4.98 (0.39) & 29.25 (1.97) & 254.13 (20.11) & 6.71 (0.50) & 968 (36) & 4203 (278) & 49390 (1877) & 962 (30) \\
        3 & MaxM-QB   & 4.68 (0.36) & \textbf{26.95 (1.73)} & 238.44 (18.39) & 6.40 (0.48) & 921 (21) & \textbf{4017 (265)} & 47009 (1104) & 922 (38) \\
        3 & Prop-QB   & \textbf{4.62 (0.35)} & 28.06 (1.82) & \textbf{235.78 (19.3)} & \textbf{6.38 (0.49)} & \textbf{918 (28)} & 4036 (279) & \textbf{46818 (1098)} & \textbf{918 (33)} \\
        \bottomrule
    \end{tabularx}
    \label{tab:queue_loss_stats}
\end{table*}

\subsection{Fairness Gains with Perimetral Queue Balancing}

We evaluate fairness based on the distribution of loss-times (across individuals and intersections) and queue lengths (across intersections of a zone), according to four different notions of fairness that are represented by four metrics: mean, maximum, sum, and standard deviation (std).
Harsanyian fairness (mean) reflects a focus on the average individual outcome.
Rawlsian fairness (maximum) reflects a focus on the worst individual outcome.
Utilitarian fairness (sum) reflects a focus on the societal rather than the individual.
Egalitarian fairness (std) reflects a focus on the equality (dispersion) of individual outcomes.


Table~\ref{tab:queue_loss_stats} reports intersections' queue lengths and loss times for the three zones.
While the gating algorithm improves efficiency, it increases average queue lengths by 5.7\% (4.62 vs. 4.37 veh uncontrolled) and maxima by 2.6\% (26.44 vs. 25.78 veh), alongside average loss times by 3.19\% (25.57 vs. 24.78 veh$\times$min) and maxima by 1.90\% (111.42 vs. 109.34 veh$\times$min); these effects persist across zones.


Perimetral queue balancing matches the gating algorithm's efficiency while improving queue and loss-time performance. 
The proportional mechanism reduces average queue lengths by 8.25\% and dispersion by 6.05\%, plus average loss times by 6.49\% and dispersion by 2.95\%. 
The max-min mechanism excels at worst-case metrics, cutting maximum queue lengths by 7.19\% and loss times by 3.73\%.
These results demonstrate the queue-homogenisation and equity benefits of the proposed mechanisms across fairness metrics.
Similar trends appear in individual trip time-loss statistics (Table~\ref{table:chabo_table}). 

\begin{table}[h]
    \centering
    \begin{tabular*}{\columnwidth}{@{\extracolsep{\fill}}lccc}
        \toprule
        \textbf{Controller} & \textbf{Mean} & \textbf{Std} & \textbf{Max} \\
        \midrule
        Uncontrolled        & 89.13 (1.34) & 209.91 (2.15) & 8313.04 (126) \\
        Gating Algorithm    & 89.62 (1.24) & 199.31 (3.48) & 8306.39 (121) \\
        Proportional QB     & \textbf{87.46 (1.08)} & \textbf{191.26 (2.87)} & 8211.66 (118) \\
        Max-Min QB          & 88.62 (1.62) & 193.17 (2.93) & \textbf{8156.43 (132)} \\
        \bottomrule
    \end{tabular*}
    \caption{\textbf{Individual TimeLoss Statistics (across scenarios).}}
    \label{table:chabo_table}
\end{table}

\FloatBarrier

\vspace{-1.5em}
\section{CONCLUSIONS}

This paper has presented a distributive perimetral queue balancing mechanism that extends classical feedback-based gating control for urban perimeter control. 
By redistributing the inflow budget according to local queue lengths at boundary intersections --via proportional (Aristotelian) or max-min (Rawlsian) allocation principles -- the approach preserves aggregate inflow targets while enhancing equity across perimetral intersections.
Microsimulation results from San Francisco's Financial District confirm conventional gating stabilises accumulations near capacity whilst improving flow and speed. 
The proposed mechanisms maintain these efficiency gains whilst further homogenising queue lengths and loss-times, and thus improving fairness grounded in different philosophical interpretations.
These findings demonstrate practical feasibility for real-time, microscopic implementations and underscore the need to integrate distributive principles into ITS for greater user acceptance and societal robustness. Future work could explore extensions to dynamic demand patterns with re-routing, and real-world field tests.

\bibliographystyle{IEEEtran}
\bibliography{references}

\end{document}